\documentclass[aps,prl,twocolumn,showpacs,groupedaddress]{revtex4}
\usepackage{graphicx}
\def\bra{\langle}
\def\ket{\rangle}

\begin{document}

\title{Phase Coherence in a Driven Double-Well System} \author{T.
Miyakawa, C.~P. Search, and P. Meystre} \affiliation{Optical
Sciences Center, The University of Arizona, Tucson, AZ 85721}

\date{\today}

\begin{abstract} We analyze the dynamics of the molecular field
incoherently pumped by the photoassociation of fermionic atoms and
coupled by quantum tunnelling in a double-well potential. The
relative phase distribution of the molecular modes in each well
and their phase coherence are shown to build up owing to
quantum mechanical fluctuations starting from the vacuum state. We
identify three qualitatively different steady-state phase
distributions, depending on the ratio of the molecule-molecule
interaction strength to interwell tunnelling, and examine the
crossover from a phase-coherent regime to a phase-incoherent
regime as this ratio increases. \end{abstract}

 \maketitle

The theoretical analysis of interacting quantum fields normally requires the
introduction of some approximation schemes that limits the detail with
which they can be understood. In the context of many-body theory, these 
approximations often take the form of a factorization (and possibly 
linearization) of the higher order correlations for the field operators such as in 
mean-field theory \cite{fetter}. There are of course exceptions to
this state of affairs, many of them in the realm of quantum
optics \cite{scully}. One such example is the micromaser, which is amenable to
an almost exact quantum description that permits a detailed
dynamical understanding of the Maxwell field inside a high-$Q$
microwave resonator \cite{micromaser_th,micromaser_ex}. 

Ultracold matter-wave fields with adjustable
interactions \cite{timmermans} are now under remarkable experimental control and are
becoming more widely available. This makes them an ideal test bed for detailed
dynamical studies of interacting quantum field theories. In this letter,
we discuss one such example, a matter-wave analog of coupled
micromasers that can be realized using experimental techniques
recently developed in the quest for molecular Bose-Einstein
condensation \cite{MBEC}. The dynamics of that system is amenable
to an essentially exact analysis that shows in particular the
build-up of phase coherence between the coupled molecular states.
As such, it sheds new light on the dynamics of quantum
fields in a system that should soon be amenable to experimental
realization.

A major current thrust in atomic, molecular and optical science is
the study of molecular Bose-Einstein condensates (BEC) formed via
either photoassociation \cite{photoassociation,wynar} or a Feshbach
resonance \cite{timmermans,wieman,jinhulet}. These efforts
have recently culminated in the realization of molecular
condensates of $^{40}$K$_2$ and $^6$Li$_2$ molecules \cite{MBEC}.

The formation of molecules by photoassociation in optical lattices
has been studied theoretically by several authors
\cite{moleculelattice}. The large energy separations that are
possible between the lowest and second Bloch band of the lattice
allow one to restrict the center-of-mass states of atoms and
molecules to the lowest energy Wannier state of each site, thereby
avoiding many of the difficulties associated with free space
\cite{goral}. Search {\it et al.}~\cite{search} recently showed
that the photoassociation of fermionic atoms into bosonic
molecules in a lattice in the Mott-insulator regime \cite{bosehubbard,SFMott}
can be mapped on to the system of the micromaser, a device that
exhibits a number of non-classical features \cite{scully}.

The present letter extends this model to include inter-well tunnelling in a double-well system
\cite{leggett,relativephase,doublewell} and analyzes the full
dynamics of the molecular field. We examine the build-up of the
relative phase between the localized molecular states of the wells
due to the combined effect of inter-well tunnelling and the
incoherent addition of molecules from photoassociation.

We identify three regimes, characterized by different orders of
magnitude of the ratio of the two-body collision strength to the inter-well
tunnelling coupling \cite{leggett}. For small values of this ratio
the relative phase of the two molecular modes becomes phase
locked at a fixed value, while for larger values that phase
remains random. This crossover of the non-equilibrium steady state
from a phase coherent regime to the phase incoherent regime is
reminiscent of the superfluid-Mott insulator transition for the
ground state in an infinite lattice \cite{bosehubbard,SFMott}.
However, in contrast to that system, this behavior occurs now in
an open system, with incoherent pumping and damping of the
matter-wave field.

We consider the two-photon stimulated Raman photoassociation 
of fermionic atoms of mass $m_f$ and spin $(\sigma=1,2)$ trapped in a
double-well potential into bosonic molecules of mass $m_b=2m_f$
\cite{photoassociation,wynar}. We assume that the system is at
zero temperature and that the number of fermions of each component
that occupy each well is less than or equal to 1 at all times. The fermions,
therefore, occupy only the lowest energy level. The molecules are
confined in the same potential and occupy only the lowest energy
level of each well.

The effective Hamiltonian for the atom-molecule system in the
lowest energy level of the wells is $ \hat{H}=\sum_{i=l,r} (
\hat{H}_{0i}+\hat{H}_{Ii})+\hat{H}_T, $ where $\hat{H}_{0i} =
\hbar(\omega_b+\delta)\hat{n}_{i}+\hbar\omega_f
(\hat{n}_{1i}+\hat{n}_{2i}) + \frac{1}{2}\hbar U_{b}
\hat{n}_{i}(\hat{n}_{i}-1)+\hbar
U_{x}\hat{n}_{i}(\hat{n}_{1i}+\hat{n}_{2i})+\hbar
U_{f}\hat{n}_{1i}\hat{n}_{2i}$, $\hat{H}_{Ii} =\hbar\chi(t)
\hat{b}^{\dagger}_i\hat{c}_{1i}\hat{c}_{2i}+h.c.$, and $\hat{H}_T =
-\hbar J_b \hat{b}^{\dagger}_l\hat{b}_r -\hbar
J_f(\hat{c}^{\dagger}_{1l}\hat{c}_{1r}
+\hat{c}^{\dagger}_{2l}\hat{c}_{2r})+h.c.$. Here $\hat{c}_{\sigma
i}$ and $\hat{b}_i$ are the annihilation operators of fermionic
atoms and bosonic molecules in the left or right wells, $i=l,r$,
respectively. The corresponding number operators $\hat{n}_{i}=
\hat{b}^{\dagger}_i\hat{b}_i$ and $\hat{n}_{\sigma
i}=\hat{c}^{\dagger}_{\sigma i}\hat{c}_{\sigma i}$ have
eigenvalues $n_{i}$ and $n_{\sigma i}$, respectively, and
$\omega_b$ and $\omega_f$ are the energies of the molecules and
atoms in the isolated wells ($J_{b,f}=0$). The terms proportional
to $U_b$, $U_x$, and $U_f$ in $\hat{H}_{0i}$ describe on-site
two-body interactions between molecules, atoms and molecules, and
atoms, respectively. The parameters $J_b$ and $J_f$ in the
tunnelling Hamiltonian, $\hat{H}_T$, are the single molecule and
atom tunnelling rates. The interaction Hamiltonian,
$\hat{H}_{Ii}$, describes the conversion of atoms into
ground-state molecules via two-photon stimulated Raman
photoassociation. The photoassociation coupling constant $\chi(t)$
is proportional to the far off-resonant two-photon Rabi frequency
associated with two nearly co-propagating lasers with frequencies
$\omega_{1}$ and $\omega_{2}$ \cite{wynar}, and $\delta$ is the
two-photon detuning between the lasers and the energy difference
of the atom pairs and molecules.

The dynamics of the molecular field in the double-well system is
governed by four mechanisms: (1) The injection of pairs of
fermionic atoms inside the double-well during  time interval $T$
when $\chi=0$, after which the atoms in each well are in the state
$|e_i\ket= \hat{c}^\dagger_{2i}\hat{c}^\dagger_{1i}|0\ket$ with
unit probability \cite{search}; (2) The unitary time evolution
of the molecular field subject to the Hamiltonian $\hat{H}_b$ during the time interval
$T$, with
    \begin{equation}
    \label{reducedH}
    \hat{H}_b=-\hbar J_b (\hat{b}_l^\dagger \hat{b}_r
    +\hat{b}_r^\dagger \hat{b}_l)
    +\hbar(U_b/4)(\hat{n}_{l}-\hat{n}_{r})^2.
    \end{equation}
    In Eq.~(\ref{reducedH}), we have neglected terms that are
functions only of $\hat{n}_l+\hat{n}_r$. This is justified as long
as the initial density matrix is diagonal in the total number of
molecules in the two wells (see below); (3) The molecular damping
at rate $\gamma$ during the intervals $T$ \cite{search}, modelled
by a master equation with a Liouvillian ${\mathcal
L}_i\hat{\rho}_b= -(\gamma/2)[\hat{b}^\dagger_i \hat{b}_i
\hat{\rho}_b-2\hat{b}_i \hat{\rho}_b
\hat{b}^\dagger_i+\hat{\rho}_b \hat{b}^\dagger_i \hat{b}_i]$ for
each well, $\hat{\rho}_b$ being the reduced density operator for
the molecules; (4) The switching on of the photoassociation
lasers in a train of short pulses of duration $\tau$ and period
$T+\tau$. We assume that $\tau$ is much shorter than all other
time scales in this model, $\tau\ll J_{b,f}^{-1},\gamma^{-1}$, so
that damping and tunnelling may be ignored while the
photoassociation fields are on. The atom-molecule conversion is
described by $F_i(\tau)\hat{\rho}_b\equiv
Tr_{a}[U_i(\tau)\hat{\rho}_{ab}(t) U^\dagger_i(\tau)]$,where
$\hat{\rho}_{ab}$ is the total density operator and $Tr_a[$ $]$
denotes the trace over the atomic variables,
$U_i(\tau)=\exp{(-i\hat{h}_i\tau/\hbar)}$ being the evolution
operator for the Jaynes-Cummings-like Hamiltonian of each well. It
describes the interaction between the molecular field and a
fictitious two-state system for fermionic atoms $|e_i \rangle $
and $|g_i \rangle = |0\rangle$ \cite{search},
\begin{eqnarray}
    \label{JCH}
    \hat{h}_i&=&\hbar\left(\omega_b+U_{x}\right)\hat{n}_{i}+
    \hbar\left(\omega_f+
    U_{x}\hat{n}_{i}\right)\hat{\sigma}_{zi}\hspace{2cm}\nonumber\\
    &+&\hbar\left(\chi(t)\hat{b}_i^{\dagger}\hat{\sigma}_{-i}+
    \chi^*(t)\hat{b}_i\hat{\sigma}_{+i} \right) +\frac{\hbar}{2}U_{b}\hat{n}_{i}(\hat{n}_{i}-1)
    \end{eqnarray}
where $\hat{\sigma}_{-i}=\hat{\sigma}_{+i}^{\dagger}=|g_i\rangle\langle e_i|$ 
and $\hat{\sigma}_{zi}=|e_i\rangle\langle e_i|-|g_i\rangle\langle g_i|$.
In Eq.(\ref{JCH}) we have dropped constant terms and made the
redefinitions $\omega_b+\delta\rightarrow \omega_b$ and
$\omega_f+U_{f}/2\rightarrow \omega_f$. For $\chi=const$,
Eq.~(\ref{JCH}) can be solved within the two-state manifolds of
each well $\{ |e_i,n_{i}\rangle , |g_i,n_{i}+1\rangle \}$.
Within each manifold, the resulting dynamics is in the form of
quantized Rabi oscillations at frequency
${\mathcal
R}_{n_{i}}=\sqrt{\left[2\omega_f-\omega_b+\left(2U_{x}-U_{b}\right)n_i\right]^2+4|\chi|^2(n_{i}+1)}$.

The photoassociation of atoms into molecules during the intervals
$\tau$ leads to the build-up of the molecular field. The
coarse-grained master equation for the reduced molecular density
operator, valid for $T\gg\tau$, is given by
    \begin{equation}
    \label{mastereq}
    \dot{\hat{\rho}}_b=\sum_{j=l,r}\left[ {\mathcal
    L}_j+T^{-1}\left\{F_j(\tau)-I_j\right\}\right]\hat{\rho}_b
    -\frac{i}{\hbar}\left[\hat{H}_b,\hat{\rho}_b\right].
    \end{equation}
The density operator for the molecular field in the number
representation for the left and right wells is given by
$\hat{\rho}_b=\sum_{n_l,n_r,m_l,m_r}\rho(n_l,n_r;m_l,m_r) |n_l\ket
|n_r\ket \bra m_r|\bra m_l|$. The initial condition for the
molecules is taken to be the vacuum state. Because the molecular
pumping and decay is the same in both wells, $\rho$ remains
diagonal in the total number of molecules in the two wells,
$N=n_l+n_r=m_l+m_r$, for all times. Consequently, the only nonzero terms
in $\rho(n_l,n_r;m_l,m_r)$ are those with $n_l+n_r=m_l+m_r$.

The master equation~(\ref{mastereq}) depends on six independent parameters:
the pump parameter, $\Theta=\sqrt{N_{ex}}|\chi|\tau$; the number of
photoassociation cycles per lifetime of the molecule, $N_{ex}=1/\gamma T$;
the two-body collision strength
and tunneling coupling strength per decay rate, $u_b=U_b/\gamma$
and $t_J=J_b/\gamma$; and finally, the
(non-)linear detuning parameters, $\eta\equiv (2\omega_f-\omega_b)/2|\chi|$
($\beta\equiv (2U_x-U_b)/2|\chi|$) from ${\mathcal R}_{n_{i}}$.
In the following, we consider for simplicity only exact resonance,
$\eta=\beta=0$, and fixed $N_{ex}=10$.
The results presented below do not depend in any qualitative manner
on the specific value of these parameters.

Introducing the angular momentum representation
$\hat{J}_{+}=\hat{J}_{-}^\dagger=\hat{J}_x+i\hat{J}_y
=\hat{b}^\dagger_l \hat{b}_r$, $\hat{J}_z=(\hat{b}_l^\dagger
\hat{b}_l -\hat{b}_r^\dagger \hat{b}_r)/2$, and
$\hat{J}^2=(\hat{N}/2)(\hat{N}/2+1)$ where
$\hat{N}=\hat{n}_l+\hat{n}_r$ is the total number operator, we
have that $\bra \hat{J}_{+} \ket =\bra\hat{J}_-\ket^*
=\sum_{n_l,n_r}\sqrt{(n_l+1)n_r}\,\rho(n_l,n_r;n_l+ 1,n_r- 1)$,
$\bra \hat{J}_z \ket = \sum_{n_l,n_r}
(n_l-n_r)\rho(n_l,n_r;n_l,n_r)/2$. Since the initial state of the
molecules in each well is the same and $\hat{H}_b$ is invariant
with respect to the interchange $l \leftrightarrow r$, the density
matrix is invariant with respect to $n_l \leftrightarrow n_r$ and
$m_l \leftrightarrow m_r$. As a result, the reduced density
matrices for the left and right wells are identical. This leads to
the same single-well molecule statistics for the two wells. The
symmetry of the density matrix with respect to the two wells
furthermore implies that $\bra \hat{J}_z\ket=\bra
\hat{J}_y\ket=0$.

\begin{figure} \includegraphics[width=8cm,height=4cm]{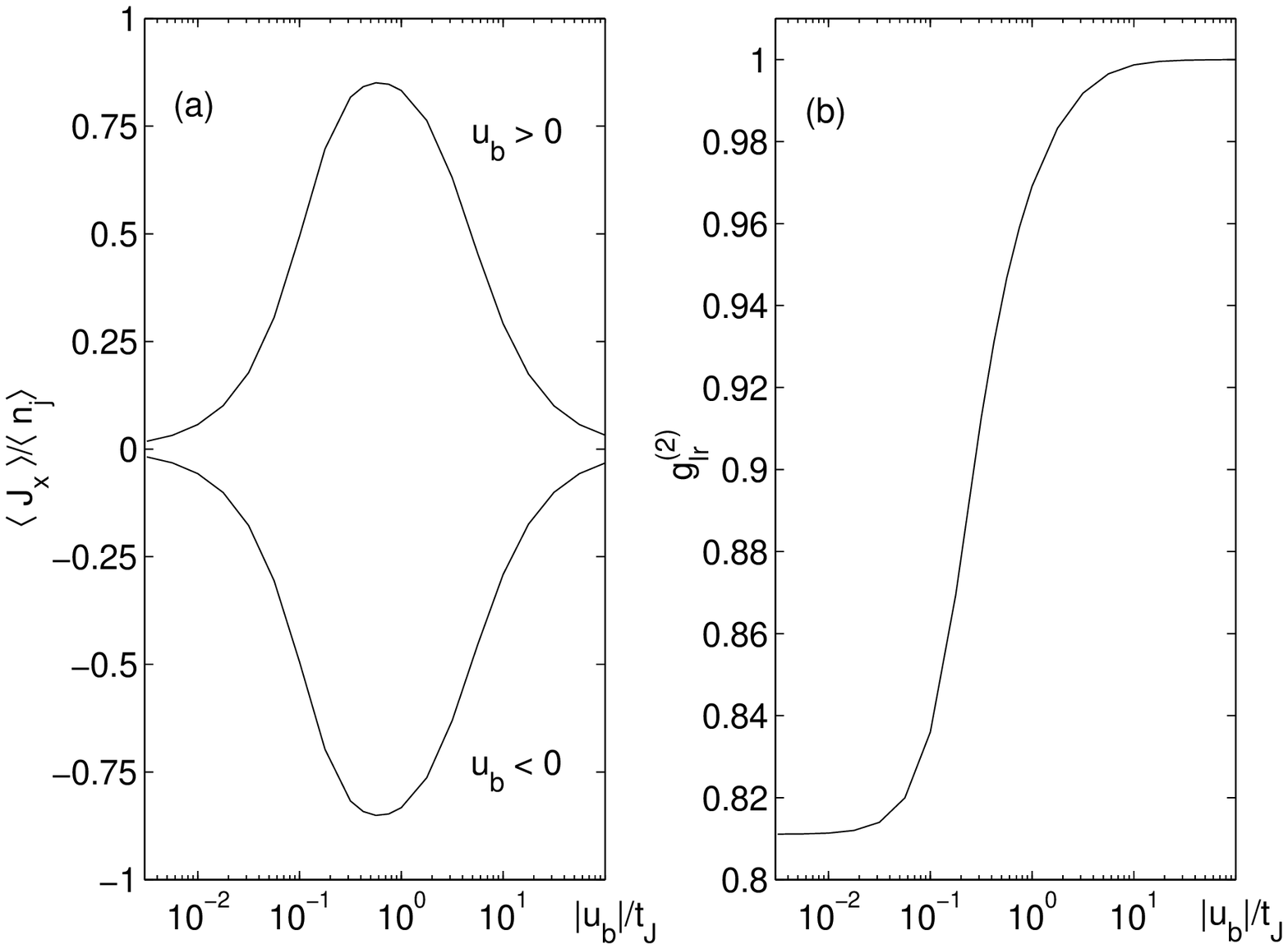}
\caption{(a) $\bra \hat{J}_x \ket/\bra n_j\ket$ and (b) $g^{(2)}_{lr}$ versus
$u_b/t_J$ for $\Theta=\pi$ and $t_J=2.5$.} \label{fig1}
\end{figure} 

Figure~\ref{fig1}(a) shows the normalized
steady-state first-order coherence $\bra\hat{J}_x\ket/\bra
\hat{n}_{j} \ket$ as a function of $u_b/t_J$ for $\Theta=\pi$ and
$t_J=2.5$. $\bra \hat{J}_x\ket$ is suppressed in both the weak and
the strong two-body coupling limits, and has an extremum at
$|u_b|/t_J\sim 0.6$. In the strong coupling regime, $|u_b|/t_J\gg
\bra \hat{N}\ket \sim 10$, the nonlinearity in $\hat{H}_b$
dominates and reduces the coherence between the localized states
of each well. We note that the average occupation numbers for each
well are relatively unaffected by $u_b/t_J$, with
$\bra\hat{n}_j\ket=\bra\hat{N}\ket/2=4.78-4.87$ for
$|u_b|/t_J=10^{2}-10^{-2.5}$.

The origin of the mutual coherence between the two molecular modes
can be determined from the equation of motion for $\bra
\hat{J}_x\ket$,
    \begin{eqnarray}
    \label{jxeq}
    \bra\dot{\hat{J}_x}\ket&=&
    \sum_{n_l,n_r}
    \sqrt{(n_l+1)n_r}\,\dot{\rho}_R(n_l,n_r;n_l+1,n_r-1)\qquad
    \nonumber\\
    &=&-U_b\bra \hat{J}_y\hat{J}_z+\hat{J}_z\hat{J}_y \ket
    -\gamma \bra \hat{J}_x \ket\nonumber\\
    &+&2T^{-1}\sum_{n_l,n_r}\sqrt{n_r}\mu_{n_l}
    \rho_R(n_l,n_r;n_l+1,n_r-1),
    \end{eqnarray}
where $\mu_{n_l}=\sqrt{n_l+1}(C^{n_l}_{n_l+1}-1)
+\sqrt{n_l+2}S^{n_l}_{n_l+1}$ with
$C^{n}_{m}=\cos{(R_n\tau/2)}\cos{(R_m\tau/2)}$,
$S^{n}_{m}=\sin{(R_n\tau/2)}\sin{(R_m\tau/2)}$, and $\rho_R$ is
the real part of $\rho$. In deriving Eq.~(\ref{jxeq}) we
used the $l\leftrightarrow r$ symmetry of $\rho$ and
$\rho(n_l,n_r;m_l,m_r)=\rho^{*}(m_l,m_r;n_l,n_r)$.

In the absence of collisions, $U_b = 0$, it can be shown from
Eq.~(\ref{mastereq}) that $\rho_R(n_l,n_r;n_l+q,n_r-q)$ with $q$
odd couple only to themselves and to the imaginary part of
$\rho(n_l,n_r;n_l+p,n_r-p)$ with $p$ even. For wells incoherently
pumped at equal rates, this implies that
$\rho_R(n_l,n_r;n_l+1,n_r-1) =0$ for all times. Alternatively,
this can be understood by noting that the expectation value
$\bra\hat{J}_x\ket$ corresponds to the difference in occupation
numbers between the in-phase,
$\hat{b}_s=(\hat{b}_l+\hat{b}_r)/\sqrt{2}$, and out-of-phase,
$\hat{b}_a=(\hat{b}_l-\hat{b}_r)/\sqrt{2}$, states of the
localized states of each well, $\hat{J}_x=\hat{b}_s^\dagger
\hat{b}_s-\hat{b}_a^\dagger\hat{b}_a$. These states are equally
populated since the bandwidth of the photoassociation pulse is larger
than their energy splitting, $1/\tau\gg J_b$. Hence
$\bra \hat{J}_x\ket=0$ for $U_b=0$, and the creation of
cross-coherence between the two wells is due solely to two-body
collisions. We also remark that a semiclassical treatment in which
$\bra\hat{J}_y\hat{J}_z+\hat{J}_z\hat{J}_y \ket$ is factorized as
$2\bra\hat{J}_y\ket\bra\hat{J}_z\ket$ \cite{doublewell} results in
$\bra \hat{J}_x\ket=0$ for all times and all values of $u_b/t_J$.
Hence, the build-up of $\bra\hat{J}_x\ket$ is a purely
quantum-mechanical effect due to quantum fluctuations.

The pump and decay mechanisms act identically on the in-phase
and out-of-phase states, hence the fact that $\bra\hat{J}_x\ket$
has the sign of $U_b$ results from the unitary time evolution from
$\hat{H}_b$, which gives
$\bra\hat{J}_x(t)\ket=(u_b/2t_J)\bra\hat{J}^2_z(t)\ket$, if
$\bra\hat{J}_x(0)\ket=\bra\hat{J}^2_z(0)\ket=0$
\cite{polkovnikov}. Since $\bra\hat{J}_z^2\ket \ge 0$, it follows that the sign of
$\bra\hat{J}_x\ket$ is determined by that of $U_b$.

Figure~\ref{fig1}(b) shows the steady-state second-order correlation function,
$g^{(2)}_{lr}=\bra\hat{n}_l\hat{n}_r\ket/\bra\hat{n}_l\ket\bra\hat{n}_r\ket$
as a function of $|u_b|/t_J$ for $\Theta=\pi$ and $t_J=2.5$. It is
related to the variance of the relative number difference, such
that $(\Delta J_z)^2=(\Delta
n_l/2)^2-\bra\hat{n}_l\ket\bra\hat{n}_r\ket
 (g^{(2)}_{lr}-1)/2+(\Delta n_r/2)^2$.
In the strong two-body coupling limit, the molecular states are
uncorrelated $g^{(2)}_{lr}=1$, and $(\Delta J_z)^2$
becomes a sum of variances for independent localized states of the
molecules.
In the weak coupling region, $|u_b|/t_J \ll \bra\hat{N}\ket^{-1}$,
the two molecular states are anti-correlated, $g^{(2)}_{lr}<1$, and
the variance $(\Delta J_{z})^2$ is greater than that of uncorrelated states.
The enhancement of relative number fluctuations indicates the locking of the
relative phase between the two wells.

To investigate the relative phase distribution of the molecular
field in the two wells, we consider the difference of the two
single-mode Pegg-Barnett phase operators \cite{peggbarnett}.
Specifically, we introduce the $(s+1)^2$ orthonormal phase states
    \begin{equation}
    |\theta_l\ket|\theta_r\ket=\frac{1}{(s+1)}\sum^s_{n_l=0}\sum^s_{n_r=0}
    e^{in_l\theta_l}e^{in_r\theta_r}|n_l\ket|n_r\ket,
    \end{equation}
where $\theta_{l}=2\pi l/(s+1)$ and $\theta_r=2\pi r/(s+1)$
($l,r=0,1,\cdots,s$) are phase variables and $s$ is a finite
value. Projection onto these phase eigenstates gives the quantum
phase distribution for the two wells,
$P(\theta_l,\theta_r)=Tr\{|\theta_l,\theta_r\ket\bra\theta_l,\theta_r|\hat{\rho_b}\}$.
Since the density matrix is diagonal in the total number of
molecules, $n_l+n_r=m_l+m_r$, the phase distribution depends only
on the relative phase, $\phi_{l,r}=\theta_l-\theta_r$, which has a
width of $4\pi$ in the phase coordinate. In order to obtain a
mod($2\pi$) distribution for the relative phase, we average over
the total phase coordinate, $\theta_l+\theta_r$, and map the
mod($4\pi$) distribution into a mod($2\pi$) one following the
method of Ref.~\cite{barnettpegg}. Using a redefined relative
phase variable, $\phi_n=2\pi n/(s+1)-\pi$ $(n=0,1,\cdots,s)$ gives
the mod($2\pi$) distribution
    $$
    P(\phi_n)=\frac{1}{s+1}\sum^{s}_{n_l,n_r=0}
    \sum^{n_r}_{k=-n_l}e^{-ik\phi_n} \rho(n_l,n_r;n_l+k,n_r-k).
    $$

\begin{figure}
    \includegraphics[width=8cm,height=8cm]{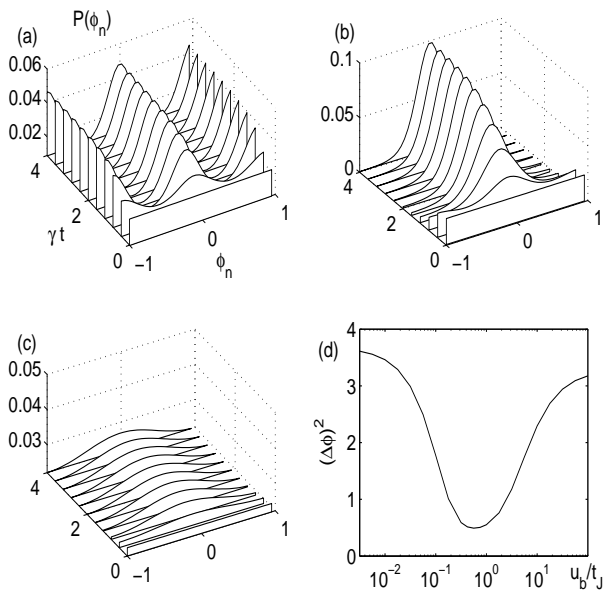}
    \caption{Time
    evolution of $P(\phi_n)$ for $\Theta=\pi$, $t_J=2.5$ and for (a)
    $u_b/t_J=0.0032$, (b) $u_b/t_J=0.5623$, (c) $u_b/t_J=56.23$; (d)
    $(\Delta \phi)^2$ of the steady state.}
    \label{fig2}
    \end{figure}

Figure~\ref{fig2} shows the evolution of the relative phase
distribution $P(\phi_n)$ in three different regimes (a)
$u_b/t_J=0.0032$, (b) $u_b/t_J=0.5623$, (c) $u_b/t_J=56.23$, for
$\Theta=\pi$, $t_J=2.5$, and $s=40$. Since the vacuum state is
taken as the initial state, the relative phase at $t=0$ is
randomly distributed, $P(\phi_n)=1/(s+1)$. For $|u_b|/t_J \alt \bra
\hat{N} \ket$, corresponding to Figs.~\ref{fig2}(a) and (b),
$P(\phi_n)$ develops a peak around $0$ and/or $\pm\pi$ in the
characteristic time $\gamma^{-1}$ needed to reach a steady state
\cite{micromaser_th}. As previously discussed, for weak two-body
interaction the in-phase and out-of-phase modes have equal
population. This leads to the bimodal phase distribution with
peaks around both $0$ and $\pm\pi$. For moderate $u_b/t_J$, the
relative phase locks in time around $0$ ($\pm\pi$), for
repulsive (attractive) two-body interactions, see
Fig.~\ref{fig2}(b). In contrast to these two regimes, when
$|u_b|/t_J\gg \bra \hat{N} \ket$, the relative phase distribution
becomes almost random, and the localized modes in the two wells
evolve independently of each other.
The steady-state phase variance, $(\Delta \phi)^2$, is shown in Fig.
~\ref{fig2}(d) as a function of the ratio $u_b/t_J$. Consistently
with Fig.~\ref{fig2}(a-c), it exhibits large fluctuations due to
the bimodal phase distribution in the case of
$u_b/t_J<\bra\hat{N}\ket^{-1}$, and becomes narrow in the region
$\bra\hat{N}\ket^{-1}<u_b/t_J<\bra\hat{N}\ket$. When
$u_b/t_J>\bra\hat{N}\ket$, the variance approaches the value
$(\Delta \phi)^2=\pi^2/3$ corresponding to a uniformly distributed
phase ($s\to\infty$).

The three regimes of phase distributions correspond to different
orders of magnitude of the ratio $u_b/t_J$, which characterizes
the behavior of the double-well system
\cite{leggett}. The crossover of the non-equilibrium steady state
from a phase-coherent regime to the random-phase situation is
reminiscent of the superfluid-Mott insulator phase transition for
the ground state of an optical lattice \cite{bosehubbard,SFMott}.
Since we consider just two sites, however, there is no sharp
transition between these regimes.

In this paper, we have studied the quantum dynamics of the bosonic
molecules in a double-well system in the presence of both
incoherent pumping and damping. We have shown that the phase
coherence of initially independent molecular states builds up due
to quantum mechanical fluctuations. We identified three
qualitatively different phase distributions for the
non-equilibrium steady states. The crossover from the phase
coherent regime to phase incoherent regime occurs as the ratio
$|u_b|/t_J$ increases, similar to the superfluid-Mott insulator
phase transition in an optical lattice.

This work is supported in part by the US Office of Naval Research,
by the National Science Foundation, by the US Army Research
Office, by the National Aeronautics and Space Administration, and
by the Joint Services Optics Program.


\end{document}